\documentstyle[preprint,revtex,12pt]{aps}
\draft
\begin{document}
\bibliographystyle{unsrt}
\begin{title}
On The Multichannel Kondo Model
\end{title}
\author{Junwu Gan}
\begin{instit}
Department of Physics, The University of British Columbia, \\
6224 Agricultural Road, Vancouver, BC~V6T~1Z1, Canada
\end{instit}
\begin{abstract}
A detailed and comprehensive study of the
one-impurity multichannel Kondo model is presented.
In the limit of a large number of
conduction electron channels $k \gg 1$, the low energy fixed point
is accessible to a  renormalization
group improved perturbative expansion in $1/k$.
This straightforward approach
enables us  to examine the scaling,
thermodynamics and dynamical response functions in great detail
and   make clear the
 following  features: i) the criticality
of the fixed point; ii) the universal non-integer degeneracy;
iii) that the compensating spin cloud has
the  spatial extent of the order of  one lattice spacing.
\end{abstract}
%
%
\vskip .5in
 Submitted to {\em J. Phys.: Cond. Matt.}
\vskip 1in



\section{Introduction}

Recently, the  non-Fermi liquid infrared fixed point of the
multichannel Kondo model, which describes a system of $k$ identical
conduction bands interacting with an impurity spin $S$,
has received considerable attention
\cite{mura86,cox87,seam91,andr91,ralp92,emer93-1,cox88,tsve90,affl91-1,affl91-2,affl91-3,sacr91-1,sacr91-2,emer92,gan93-1,affl93,cox93,emer93-2}.
The interest is aroused by
its potential application in
 the metallic glasses \cite{mura86,ralp92,sacr91-2},
and heavy fermion Uranium alloys \cite{cox87,seam91,andr91}
 where more examples have been
found with their low temperature behavior falling out of the
usual Fermi liquid expectation \cite{nozi74,lee86}.
It is enhanced by
the  more interesting observation of non-Fermi liquid
behavior in the normal state of
the high Tc cuprates \cite{ande92,varm89} and its resemblance
 to the low energy
behavior of the multichannel Kondo model \cite{emer93-1,gan93-1}.

Although this model is more than ten years old \cite{nozi80}
 and has been attacked by various
 methods \cite{affl91-1,emer92,gan93-1,cox93,crag80,andr84,tsve84},
there is still room left
for  a simple interpretation(or reinterpretation) of
the physics of the low energy fixed point
  which is often said to be
nontrivial.
This task is easily achieved in the limit
 $k \gg 1$ where the low energy fixed point has
  a value  of
order $1/k$  for the coupling constant and
is accessible to a renormalization
group(RG) improved perturbative expansion in $1/k$.
Some physical quantities difficult to calculate by the other methods
are readily obtained by this perturbative approach and the results
provide new  insight into the fixed point.
That the nature of the fixed point remains qualitatively the
same when continuing
$k$ to small values as long as $k > 2S$
has been demonstrated by the results
of Bethe Ansatz or conformal field  methods \cite{affl91-1,andr84,tsve84}.
In this paper, we calculate a long list of physical
quantities
to  the leading or sub-leading order in  $1/k$.
A simple physical picture can be sketched based on
 these results and  the previous understanding.

For  an antiferromagnetic  Kondo interaction between the the impurity
spin and the conduction electrons, the impurity spin pulls in
conduction electrons with opposite spin. This increases the
spin-flip exchange which in turn enhances the attraction
of the conduction electrons with opposite spin to the impurity spin.
The result of this cooperative enhancement of the Kondo interaction
is that one conduction electron from each of the $k$ channels
is pulled in to screen the impurity spin.
Since $k > 2S$, the conduction electrons overscreen the impurity
spin. Additional conduction electrons must come in to remedy
the over-performance. The process keeps going on resulting
 in a critical system.
It is important to realize that
 the system is only critical along the time axis.
As short time details are averaged out, the system approaches
a universal limiting behavior
 given by the infrared fixed point.
The impurity spin is  asymptotically screened in the
ground state  which is a spin singlet.
However, the infrared fixed point can be reached only in an
infinite system in which the energy levels form a continuum.
In finite systems, the  typical distance between
discrete energy levels, of order $1/L$
for a system of linear dimension $L$, cuts off the scaling
toward the fixed point.
The ground state of a finite system has a residual spin and is
doubly degenerated.
A common definition of the residual spin is
 $\sqrt{T \chi(T)}$ as the temperature $T \rightarrow 0$,
where $\chi(T)$ is the magnetic susceptibility.
As we shall see, it is nonzero for a finite system.
We shall also show that the entropy as given by the coefficient
of the linear $T$ term in the free energy is indeed $\ln (2S+1)$
for a finite system.

An intriguing feature of the multichannel Kondo model is
that the entropy of an infinite system reduces to a
smaller universal value \cite{tsve84,affl91-2,gan93-1}
 although the ground state is a spin singlet.
If  the entropy  is still defined as the logarithm of
the ground state degeneracy,  the ground
state could have a universal non-integer degeneracy.
Since the entropy is deduced by
calculating the free energy at small but finite temperature $T$
 which cuts off the scaling  toward the fixed point,
 it is  natural to relate the entropy to the effective
residual spin at $T$. A necessary implication of
this relation  is that a tiny magnetic field
will lift the degeneracy at $T \rightarrow 0$
as we shall see later.

Another interesting issue pertaining to all kind of Kondo problems,
either overscreened or exactly screened,
is the spatial size of the conduction electron screening cloud.
This is a point underlying  the resonant level nature of
 the   Kondo problems. Considering the one-channel Kondo problem,
it is well known that below the Kondo temperature $T_{K}$
the effective Kondo interaction enters the strong coupling regime.
The energy gain from screening the impurity spin is of
order $T_{K}$. If the screening cloud formed a localized
bound state with the impurity spin, it would have
a spatial spreading of $v_{F}/T_{K} \gg 1/k_{F}$ \cite{nozi85}.
NMR experiment ruled out any conduction electron
screening cloud bigger than one lattice spacing \cite{boyc74}.
Although the physical argument for the screening cloud
to have a size of $1/k_{F}$ has been
given \cite{ande84,varm87}, here we  calculate the Knight
shift for the multichannel Kondo model and
explicitly show that the only length scale is $1/k_{F}$.
We also explain why the same conclusion can be extended to
the exactly screened case(k=1).

The  paper is organized as follows.
In section~\ref{PT}, the multichannel Hamiltonian and the
Popov method are briefly recaptured.
In section~\ref{CESE}, the general
integral expression
for the conduction electron self-energy to the order
${\cal O}(k^{-4})$ is derived
which  can serve as a future
reference. In section~\ref{RGEaS},  the integrals in
the self-energy  are evaluated  at $T=0$ and the results are
used to derive the RG equation and running coupling constant.
In section~\ref{SRaR},  the scaling solutions
for the conduction electron scattering rate and resistivity
are obtained.
The free energy is calculated in section~\ref{SHaE}
from which the specific heat and entropy are deduced.
The magnetic susceptibility and field dependent magnetization
are calculated in section~\ref{FDM} for an equal
spin gyromagnetic ratio for the
conduction electron and impurity spin.
The dynamical susceptibility for the impurity spin
is calculated in
section~\ref{DS} and  the relaxation rate is deduced.
The general case with different gyromagnetic ratios
is considered in section~\ref{MSF}.
The Knight shift in the space surrounding the impurity spin
is calculated in section~\ref{KS}.
The last section is devoted to a discussion of related issues.
Part of the results has been briefly reported in \cite{gan93-1}.
Some details for the dynamical spin correlation functions
are included in the appendix~\ref{dyn-correl}.

\section{Popov Technique  }		\label{PT}

Without losing generality, we consider the local impurity to be
a spin $S=1/2$.
The multichannel Kondo Hamiltonian in the magnetic field is
\begin{equation}
H= \sum_{ \stackrel{\vec{k},\mu=\pm}{\lambda=1 \cdots k} }
( \epsilon_{\vec{k}} + \mu h)
c^{\dagger}_{\vec{k}\mu\lambda} c_{\vec{k}\mu\lambda}
+  2  \mu_{S} \vec{h} \cdot \vec{S}
+\frac{J}{\cal N} \sum_{\vec{k},\vec{k'}}
\sum_{ \stackrel{ \mu,\nu=\pm   }{	\lambda=1 \cdots k}	}
c^{\dagger}_{\vec{k}\mu\lambda} \frac{\vec{\sigma}_{\mu\nu}}{2}
c_{\vec{k}'\nu\lambda} \cdot \vec{S},
\label{hamil1}
\end{equation}
where $\vec{S}$ is a spin-1/2 operator,
$\vec{\sigma}_{\mu\nu}$ are
the Pauli matrices,
$J$ is the Kondo interaction strength
 and ${\cal N}$ is the number of
lattice sites. We have
set the  conduction electron gyromagnetic ratio and Bohr magneton
equal to one so that the magnetic field $h$ has the dimension of energy.
We have introduced a parameter $\mu_{S}$
to account for
possible difference
between the conduction electron
and  impurity spin: $\mu_{S}$ is
equal to the gyromagnetic ratio  of the impurity spin
divided by that of the
conduction electron.
We shall adopt the usual cutoff scheme for the  conduction electron band:
$-D< \epsilon_{\vec{k}} <D$,  with  a constant density of states
$\rho$ per spin per channel.

Representing the impurity spin in term of pseudofermions
\[ 	\vec{S}= \frac{1}{2} \sum_{\mu,\nu=\pm} f^{\dagger}_{\mu}
\vec{\sigma}_{\mu\nu} f_{\nu}	,	\]
Popov made an observation
that the Kondo Hamiltonian without imposing constraint on the
pseudofermions is disconnected in the pseudofermion
charge sectors. That is
\[ {\cal Z} =  {\rm Tr} \, e^{-\beta H} =
{\cal Z}_{0} + {\cal Z}_{1} +
 {\cal Z}_{2} =
{\rm Tr} \delta(\hat{n}_{f}) e^{-\beta H} +
{\rm Tr}   \delta(\hat{n}_{f}-1) e^{-\beta H} +
{\rm Tr}  \delta(\hat{n}_{f}-2)  e^{-\beta H}	.	\]
Among these three separate contributions,
only  ${\cal Z}_{1} $ from the subspace
\[ 	\hat{n}_{f} = \sum_{\sigma=\pm} f_{\sigma}^{\dagger} f_{\sigma}
			=1			 \]
is physical.
Popov's technique \cite{popo88}
 is to add an imaginary chemical potential,
$i  \omega_{0}= i \pi/(2\beta) $,
to the pseudofermions so that $ {\cal Z}_{0} + {\cal Z}_{2} 	=0$.
Using this trick,
the partition function in the path integral formalism
can be represented  as
\begin{eqnarray}
{\cal Z} &=& \int {\cal D}[\bar{c},c,\bar{f},f]
\exp\left[ -\int^{\beta}_{0} d\tau ({\cal L}_{0}+H) \right] ,  \\
{\cal L}_{0} &=&  \sum_{\vec{k},\mu, \lambda}
\bar{c}_{\vec{k}\mu\lambda} \partial_{\tau} c_{\vec{k}\mu\lambda}
+\sum_{\mu} \bar{f}_{\mu} (\partial_{\tau}+ i\omega_{0}) f_{\mu} .
\end{eqnarray}
There is no additional constraint.
The standard
perturbation method then follows from this path integral representation.
 The impurities are assumed
to be randomly distributed in space. The averaging over the impurity
distribution is done according to the standard recipe
as in the case of the spinless impurities \cite{mahan90,abri65} and
we only keep  contributions linear in the impurity density $n_{i}$.
The Feynman rules for constructing diagrams are listed
in the appendix~\ref{feyn-rule}.

\section{Conduction Electron Self-Energy}	\label{CESE}

The first thing we shall
 calculate is the  conduction electron
self-energy in the absence of the
external magnetic field.  Up to the
order $J^{4}$, $J^{5}k$ and $J^{6}k^{2}$, the relevant diagrams
are given in Figure~\ref{self-en-fig}. The two subscripts
in each term of the self-energy expansion
indicate the powers of $J$ and $k$ respectively.
After lengthy algebra, the final results at finite
temperature are
\begin{eqnarray}
\Sigma(i\omega_{n},T) &=& \Sigma_{(2,0)} +
\Sigma_{(3,0)} + \Sigma_{(4,1)}
+ \Sigma_{(4,0)} + \Sigma_{(5,1)}
+ \Sigma_{(6,2)}	, 		\\
\Sigma_{(2,0)}(i\omega_{n}) &=& -\frac{3n_{i}}{16} J^{2}\rho
\int \frac{d \epsilon}{i\omega_{n} -\epsilon} 	,
			\label{sigma20}				  \\
\Sigma_{(3,0)} (i\omega_{n}) &=& \frac{3n_{i}}{16} J^{3}\rho^{2}
\int d\epsilon \, d\epsilon_{1}
\frac{ \tanh\left( \frac{\beta \epsilon_{1} }{2} \right) }
{(i\omega_{n} -\epsilon)(\epsilon -\epsilon_{1}) }  , 		  \\
\Sigma_{(4,1)} (i\omega_{n}) &=& \frac{3n_{i}}{16} J^{4}\rho^{3} k
\int  \frac{ d\epsilon \, d\epsilon_{1} \, d\epsilon_{2} \;
 \varphi^{\epsilon}_{\epsilon_{1} \epsilon_{2} }	}
{ (\epsilon -\epsilon_{1})
(i\omega_{n} -\epsilon_{2})
(i\omega_{n} -\epsilon_{2}+\epsilon -\epsilon_{1}) }   ,	 \\
\Sigma_{(4,0)} (i\omega_{n}) &=& \frac{9 n_{i}}{32} J^{4} \rho^{3}
 \int  \frac{ d\epsilon \, d\epsilon_{1} \, d\epsilon_{2} }
{(i\omega_{n} -\epsilon)(\epsilon -\epsilon_{1})
(\epsilon -\epsilon_{2}) }
 \left[ \varphi^{\epsilon_{1}}_{\epsilon_{2}}  - \frac{3}{8} \right]	,
			\label{sigma40}
\end{eqnarray}
\begin{eqnarray}
  \Sigma_{(5,1)} (i\omega_{n}) &=& - \frac{3 n_{i}}{32} J^{5} \rho^{4} k
 \int  \frac{  d\epsilon_{1} \,
d\epsilon_{2} \, d\epsilon_{3} \, d\epsilon_{4} \;  \;
\xi^{\epsilon_{1}}_{\epsilon_{2} \epsilon_{3} } }
{ (\epsilon_{1} - \epsilon_{2} )   (\epsilon_{4} - \epsilon_{3} )
 (\epsilon_{1} - \epsilon_{2} + \epsilon_{4} - \epsilon_{3} )
(i\omega_{n} - \epsilon_{4}) }
			\nonumber		\\
    &- &   \frac{3 n_{i}}{32} J^{5} \rho^{4} k
 \int  \frac{   d\epsilon_{1} \,
d\epsilon_{2} \, d\epsilon_{3} \, d\epsilon_{4} \;
 \tanh\left( \frac{\beta \epsilon_{3} }{2} \right)  \,
 \varphi^{\epsilon_{2}}_{\epsilon_{4} \epsilon_{1}} }
{  (\epsilon_{1} - \epsilon_{2} )   (\epsilon_{4} - \epsilon_{3} )
 (\epsilon_{1} - \epsilon_{2} + \epsilon_{4} - \epsilon_{3} )
 (i\omega_{n} - \epsilon_{4} + \epsilon_{2} - \epsilon_{1} ) }
		\nonumber		\\
      &+&   \frac{3 n_{i}}{16} J^{5} \rho^{4} k
 \int  \frac{   d\epsilon_{1} \,
d\epsilon_{2} \, d\epsilon_{3} \, d\epsilon_{4} \;
\tanh\left( \frac{\beta \epsilon_{3} }{2} \right)   \,
 \varphi^{\epsilon_{2}}_{\epsilon_{4} \epsilon_{1}}	}
{ (\epsilon_{1} - \epsilon_{2} )
(i\omega_{n} - \epsilon_{4})
(i\omega_{n} - \epsilon_{4} + \epsilon_{2} - \epsilon_{1} ) }
\left( \frac{1}{ \epsilon_{4} - \epsilon_{3}}
+\frac{1}{  \epsilon_{2} - \epsilon_{3}  } \right)
		\nonumber			\\
 &-& \frac{21 n_{i}}{128} J^{5} \rho^{4} k
 \int  \frac{   d\epsilon_{1} \,
d\epsilon_{2} \, d\epsilon_{3} \, d\epsilon_{4} \;
\tanh\left( \frac{\beta \epsilon_{3} }{2} \right)   \,
 \varphi^{\epsilon_{2}}_{\epsilon_{4} \epsilon_{1}}	}
{ (\epsilon_{1} - \epsilon_{2} )  (\epsilon_{1} - \epsilon_{3} )
(\epsilon_{2} - \epsilon_{3} )
(i\omega_{n} - \epsilon_{4} +  \epsilon_{2} - \epsilon_{1} )   }	,
		\label{sigma51}
\end{eqnarray}
\begin{eqnarray}
     \Sigma_{(6,2)} (i\omega_{n}) &=&
  -\frac{3 n_{i} }{32} J^{6} \rho^{5} k^{2} \int
\frac{  d\epsilon_{1} \, d\epsilon_{2} \, d\epsilon_{3}
\, d\epsilon_{4} \, d\epsilon_{5} }{ (i\omega_{n} - \epsilon_{5})
 (\epsilon_{1} - \epsilon_{2} )    (\epsilon_{3} - \epsilon_{4} ) }
			\nonumber		 \\
  & \times & \left[  \frac{   \varphi^{\epsilon_{1}}_{\epsilon_{2}}
\;  \varphi^{\epsilon_{3}}_{\epsilon_{4}\epsilon_{5}}	  }{
(i\omega_{n} - \epsilon_{5} + \epsilon_{3} - \epsilon_{4} )
(\epsilon_{1} - \epsilon_{2} + \epsilon_{3} - \epsilon_{4} ) 	  }
  +   \frac{   \varphi^{\epsilon_{1} \epsilon_{3}}_{\epsilon_{2}
	\epsilon_{4} \epsilon_{5} }  }{
i\omega_{n} - \epsilon_{5}
+ \epsilon_{1} - \epsilon_{2} + \epsilon_{3} - \epsilon_{4}  }
	\right.		\nonumber 	\\
 & \times &  \left.
 \left(
\frac{ 2 }{  i\omega_{n} - \epsilon_{5} + \epsilon_{3} - \epsilon_{4} }
+ \frac{1}{  \epsilon_{1} - \epsilon_{2} + \epsilon_{3} - \epsilon_{4} }
		\right) 	\right]  ,
		\label{sigma62}
\end{eqnarray}
where  the integration range is always $[-D,D]$.
Whenever necessary,
the above integrals(and integrals throughout this paper)  take
principal value. We have introduced
following shorthand notations,
\begin{eqnarray*}
 & & \varphi^{\epsilon_{1}}_{\epsilon_{2}} =
f(\epsilon_{1}) \,f(-\epsilon_{2})
+ f(-\epsilon_{1}) \,f(\epsilon_{2})  ,
					\\
 & & \varphi^{\epsilon_{3}}_{\epsilon_{4}\epsilon_{5}} =
 f(\epsilon_{3}) \,f(-\epsilon_{4})
 \,f(-\epsilon_{5}) +
f(-\epsilon_{3}) \,f(\epsilon_{4})  \,f(\epsilon_{5}) ,
					\\
 & & \xi^{\epsilon_{1}}_{\epsilon_{2}\epsilon_{3}} =
  f(\epsilon_{1}) \,f(-\epsilon_{2})
 \,f(-\epsilon_{3}) -
f(-\epsilon_{1}) \,f(\epsilon_{2})  \,f(\epsilon_{3}) ,
					\\
  & & \varphi^{\epsilon_{1} \epsilon_{3}}_{\epsilon_{2}
\epsilon_{4} \epsilon_{5} }  =
 f(\epsilon_{1}) f(-\epsilon_{2}) f(\epsilon_{3})
 f(-\epsilon_{4}) f(-\epsilon_{5})
			 +  f(-\epsilon_{1}) f(\epsilon_{2})
 f(-\epsilon_{3})
 f(\epsilon_{4}) f(\epsilon_{5}) ,
\end{eqnarray*}
where $f(\epsilon)=1/(e^{\beta \epsilon} +1 )$, is the Fermi-Dirac function.

\section{Renormalization Group Equation and Solution} \label{RGEaS}

Under the analytic continuation $i\omega_{n} \rightarrow \omega + i 0^{+}$,
$\Sigma(\omega + i 0^{+}) = \Sigma'(\omega) + i  \Sigma''(\omega) $.
The imaginary part of the self-energy
$\Sigma''(\omega)$  is proportional
to the total scattering rate of the conduction electron of
energy $\omega$.
 At $T=0$, the integrations in the imaginary part of the
self-energy (\ref{sigma20})-(\ref{sigma62})
can be carried out and  we obtain
\[
\Sigma''(\omega,D,g) = \frac{3\pi n_{i}}{16 \rho}
g^{2} \left( P_{0} +P_{1} g \ln\widetilde{\omega}
+P_{2} g^{2}k\ln\widetilde{\omega} +P_{3} g^{2}k
+P_{4} g^{2} \ln^{2}\widetilde{\omega}
	 + P_{5} g^{3}k \ln^{2}\widetilde{\omega}
			\right.     \]
\begin{equation}
 \hspace{.4in}  \left.
+P_{6} g^{3}k \ln\widetilde{\omega} + P_{7} g^{3}k
+P_{8} g^{4}k^{2} \ln^{2}\widetilde{\omega}
+ P_{9}g^{4} k^{2} \ln\widetilde{\omega} + P_{10} g^{4} k^{2} \right)  ,
			\label{rate}
\end{equation}
where $g=J\rho$ and
$\widetilde{\omega}=\omega/D$.
All coefficients $P_{0}$ to $P_{10}$ are known:
 $P_{0}=1$, $P_{1}=-2$,
$P_{2}=1$, $P_{3}=\ln 2 -1$, $P_{4}=3$, $P_{5}=-7/2$, $P_{6}=5-3\ln 2$,
$P_{8}=1$, $P_{9}=2\ln 2 -5/2$. The coefficients $P_{7}$
and $P_{10}$ are  not needed for deriving the
RG equation to the sub-leading order.
But they can be found from
(\ref{sigma51}) and (\ref{sigma62}).
Note that the $g^{4}$ order contributions from (\ref{sigma40})
contain only a $\ln^{2}(\widetilde{\omega})$ term.
Since we expect the fixed point $g^{*}$ to be of order $1/k$ \cite{nozi80},
our result~(\ref{rate}) includes
all contributions up to the order ${\cal O}(k^{-4})$.

Since the dimensionless scattering rate  (\ref{rate}) must be invariant
under RG transformation, we obtain the
following equation,
\begin{equation}
\left(\frac{\partial}{\partial \ln D}
+ \beta(g) \frac{\partial}{\partial g} \right)
\rho\Sigma''(\omega,D,g) = 0   .
		\label{rgeqn}
\end{equation}
The beta-function, $\beta(g)$,  of the Kondo interaction $g$
can be deduced from (\ref{rate})
using the standard RG technique \cite{coll84}.
The equation~(\ref{rgeqn}) can be regarded as an equation which generates
logarithmic series
with fixed $g$,
\begin{equation}
\Sigma''(\omega,D,g) = - \beta(g)  \frac{\partial}{\partial g}
\int d(\ln D) \Sigma''(\omega,D,g) + {\rm const.}
		\label{pseries}
\end{equation}
Substituting (\ref{rate}) into (\ref{pseries}), the integration over
$\ln D$ can be carried out easily.
Expressing $\beta(g)$ to the sub-leading order as
\begin{equation}
\beta(g)=g^{2} ( \kappa_{1} + \kappa_{2} gk + \kappa_{3} g
+ \kappa_{4} g^{2}k + \kappa_{5} g^{3}k^{2})
\end{equation}
and substituting it into  (\ref{pseries}), we obtain following
equations for  the coefficients $\kappa_{1}$  to $\kappa_{5}$ from
 (\ref{pseries}) by equating order by order
the coefficients of the polynomials on the left and right sides,
\begin{eqnarray}
\kappa_{1} &=& \frac{P_{1}}{2P_{0}} = -1    	,		\\
\kappa_{2} &=& \frac{P_{2}}{2P_{0}} =\frac{1}{2} ,		\\
\kappa_{3} &=& 0	,			\\
\kappa_{4} &=& \frac{P_{6}- 4P_{3} \kappa_{1}}{2P_{0}}
			=\frac{1+\ln 2}{2} , 		\\
\kappa_{5} &=& \frac{P_{9}- 4P_{3} \kappa_{2}}{2P_{0}} = -\frac{1}{4}  .
\end{eqnarray}
There are three consistency equations
\begin{eqnarray*}
P_{4} &=& \frac{3}{2} P_{1} \kappa_{1}  , 		\\
P_{5} &=& 2P_{2} \kappa_{1} + \frac{3}{2} P_{1} \kappa_{2} ,  	\\
P_{8} &=& 2P_{2}\kappa_{2}  ,
\end{eqnarray*}
which are all satisfied.

The obtained beta-function can be written explicitly as
\begin{equation}
\beta(g) = -g^{2} +\frac{k}{2} g^{3}
+\frac{k}{2}(1+\ln 2) g^{4} - \frac{k^{2}}{4} g^{5} .
\label{beta}
\end{equation}
The intermediate fixed point $g^{*}$,  determined
by $\beta(g^{*})=0$, and the slope of the $\beta(g)$
at the fixed point are then easily found to be
\begin{eqnarray}
g^{*} &=& \frac{2}{k} \left(1- \frac{2\ln 2}{k} \right) ,  \\
 \Delta  &=&  \beta'(g^{*})  \;
= \;  \frac{2}{k} \left(1-\frac{2}{k} \right) .
		 \label{slop}
\end{eqnarray}
We find indeed $g^{*} \sim 1/k$ thus allowing a reliable expansion
in the $k \rightarrow \infty$ limit.
The slope which determines the
critical exponent is universal.
The perturbative result (\ref{slop}) agrees with
the conformal field result $2/(k+2)$ \cite{affl91-1} up to the
sub-leading order in the $1/k$ expansion.

The running coupling constant $g_{R}(\omega)$ is
determined by the differential equation
\begin{equation}
\frac{ d \, g_{R} }{ d\ln \omega } = \beta(g_{R}) ,
	\label{grun-eq}
\end{equation}
 with the initial condition
$g_{R}(\omega\!=\!D)=g$.
 With our perturbative
$\beta(g)$ of (\ref{beta}), the solution of (\ref{grun-eq})
covering the full
 range of energy scale
from $\omega \ll T_{K}$ to $\omega \gg T_{K}$
can be obtained.
We rewrite (\ref{grun-eq}) in the integral form,
\[ \int^{\omega}_{D} d \ln \omega' =
\int^{g_{R}(\omega)}_{g} \frac{ d g'}{\beta(g')}
\simeq g^{*} \int^{g_{R}(\omega)}_{g}
 \frac{ d g'}{ (g')^{2} (g'-g^{*}) }
\left[ 1 + g' \ln 2 - \frac{k}{2} (g')^{2} \right]^{-1} .  \]
This leads to the full solution
\begin{equation}
 | g_{R}(\omega) - g^{*} | = |g-g^{*}|
 \left( \frac{\omega}{T_{K}} \right)^{\Delta} \;
\left[ g_{R}(\omega) \right]^{k\Delta/2 } \;
e^{ - \Delta/ g_{R}(\omega)   } ,		\label{grun-ful}
\end{equation}
where  $T_{K}=Dg^{k/2}\exp(-1/g)$.
At $\omega < T_{K}$, it has an asymptotic form of,
\begin{equation}
g_{R}(\omega)=g^{*} - \zeta \left(\frac{\omega}{T_{K}}\right)^{\Delta}  ,
	\hspace{.3in}
 \zeta = (g^{*}-g) \, (g^{*})^{k\Delta/2 }  \,
e^{- \Delta/g^{*}  }   .	\label{grun}
\end{equation}
For an  initial condition of weak coupling, $g \rightarrow 0$
and $D \rightarrow \infty$,
$\zeta= (g^{*})^{1+k\Delta/2}\, e^{-\Delta/g^{*}}$.
{}From (\ref{grun-ful}), it is interesting to note that the running coupling
constant has a power law behavior not only at low
energy $\omega < T_{K}$ but also at high energy $\omega > T_{K}$,
underlying the critical nature of the system.

\section{Scattering Rate and Resistivity} 	\label{SRaR}

To obtain the scaling solution for
the conduction electron scattering rate
$\rho \Sigma''$, we can use the
RG invariance,
\begin{equation}
\rho \Sigma''(\omega,D,g_{R}(D)\!=\!g)
=\rho \Sigma''(\omega,D',g_{R}(D'))  .
\end{equation}
Choosing $D'=\omega$,  the logarithmic terms in
$\rho \Sigma''(\omega,\omega,g_{R}(\omega))$ of  (\ref{rate}) drop out and
the scaling form for (\ref{rate}) is
\begin{equation}
\rho\Sigma''\left(\omega,D,g \right)
 = \frac{3\pi n_{i}}{16}
\left[ g_{R}^{2}(\omega)-(1-\ln 2)k g_{R}^{4}(\omega) \right]
\simeq \frac{3\pi n_{i}}{4(k+2)^{2}} \left[ 1 -
A_{0} \left(\frac{\omega}{T_{K}}\right)^{\Delta}   \right]   ,
		\label{rate-f}
\end{equation}
where  $A_{0}=k\zeta [1- (4-6\ln 2)/k ] $.
A similar RG procedure will be performed repeatedly later.
Note that instead of a Lorentzian frequency dependence
in the exactly screened Kondo problem,
$\rho \Sigma'' \sim T_{K}^{2}/(\omega^{2}+T_{K}^{2})$,
the scattering rate~(\ref{rate-f}) has a cusp at $\omega=0$.
 At nonzero temperature, the cusp is expected to be
smoothed out.

In order to calculate the  resistivity, we need the temperature
dependent imaginary part of the conduction electron  self-energy
$\Sigma''(\omega,T)$.
Identifying the transport relaxation time  due to the Kondo exchange as
$\tau_{\rm ex}(\omega,T) =  1/[2\Sigma''(\omega,T)] $
since there is only $s$-wave scattering \cite{mahan90},
the total scattering rate is
 $1/\tau(\omega,T)=1/\tau_{0} + 1/\tau_{\rm ex}(\omega,T)$,
where $\tau_{0}$ is the ordinary relaxation time in the absence
of  the Kondo interaction.
The ordinary scattering $1/\tau_{0}$ could arise from
spinless impurities or defects which are assumed to be located
at different lattice sites and uncorrelated
to the impurity spins. A low impurity spin density is assumed,
$n_{i} \ll 1$, such that $\tau_{0} \ll \tau_{\rm ex}$.
The total relaxation
time is substituted into the conductivity
expression \cite{abri65},
\begin{equation}
\sigma(T)=\frac{ n_{e} e^{2}}{m_{e}} \int^{\infty}_{0} d\omega
\frac{ \tau(\omega) }{ 2T \cosh^{2} \left(\frac{\omega}{2T}\right) }
\simeq \frac{ n_{e} e^{2}}{m_{e}} \int^{\infty}_{0} d\omega
\frac{ \tau_{0} }{ 2T \cosh^{2} \left(\frac{\omega}{2T}\right) }
\left( 1 - \frac{\tau_{0}}{\tau_{\rm ex}(\omega)} \right)  ,
\end{equation}
with $e$, $m_{e}$ and $n_{e}$
denoting the conduction electron charge, mass and  density
 respectively.
With  our perturbative expression for
 $\Sigma''(\omega,T)$ to the order $k^{-3}$,
we find for the resistivity due to  Kondo scattering,
\begin{eqnarray}
 \delta \rho_{e}(T) &=& \frac{3\pi}{16} \frac{m_{e}}{n_{e}e^{2} \rho}
n_{i} g^{2} \int^{\infty}_{0} \frac{ dx}{\cosh^{2}\frac{x}{2} }
\left[ 1 - g \int^{D/T}_{-D/T} dy \frac{\tanh\frac{y}{2}}{x-y}
-g^{2}k \; \Psi\left(x,\frac{D}{T}\right) \right]
			\nonumber 	\\
 &=& \frac{3\pi}{8} \frac{m_{e}}{n_{e}e^{2} \rho}
n_{i} g^{2} \left[ 1 + {\cal O}(g, g^{2}k)
 + {\cal O}(g, g^{2}k) \times \ln\frac{D}{T} \right] ,
		\label{resis-exp}
\end{eqnarray}
where
\begin{equation}
\Psi\left(x,\frac{D}{T}\right) =
\int^{D/T}_{-D/T} dy \int^{D/T}_{-D/T} dy_{1}
\frac{ \frac{1}{e^{y}+1}  \frac{1}{e^{-y_{1}}+1} }{(y-y_{1})^{2}}
\left( 1 - \frac{1}{e^{y_{1}-y-x}+1} - \frac{1}{e^{y_{1}-y+x}+1} \right) .
\end{equation}
Some difficult integrals
have to be evaluated to obtain
the sub-leading terms in (\ref{resis-exp}).
So we limit ourselves to the leading order.
Performing the same RG procedure as  (\ref{rate-f}),
we obtain
\begin{equation}
\delta \rho_{e} = \frac{3\pi^{2} }{4 k^{2}} \frac{n_{i}}{n_{e}}
\rho_{e}^{(0)}
\left[ 1- k\zeta \left(\frac{T}{T_{K}}\right)^{\Delta} \right]    ,
			\label{resist}
\end{equation}
where $\rho_{e}^{(0)}=4\pi/(e^{2}k_{F})$, is the
resistivity in the unitary limit and
  $k_{F}$ denotes the  Fermi wave vector which is related
to the conduction electron density of states
through $\rho=k_{F}m_{e}/(2\pi^{2})$.
The $T=0$ value of the resistivity and the exponent  $\Delta$ have
been reported \cite{affl91-3} and
an exact expression for the resistivity for $k=2$
up to a constant factor $\zeta$ has been
derived recently \cite{affl93}.
 As in the exactly screened Kondo problem, the corresponding
resistivity decreases upon increasing temperature.

An important point is that even at $T=0$, there is still inelastic
scattering. In other words, the impurity spin together
with its asymptotical
screening cloud(see magnetization section) cannot be viewed
as inert. Let's see how the assumption of elastic scattering
only at $T=0$ would run into trouble.
Following \cite{nozi74},
the total resistivity can be divided into
elastic and inelastic contributions,
\begin{equation}
 \delta\rho_{e} = \delta\rho_{e}^{\rm el}
+ \cos(2\phi_{0}) \; \delta\rho_{e}^{\rm in} ,   \label{resis-sum}
\end{equation}
where  $\phi_{0}$ is the  scattering
phase shift at the Fermi energy.
This separation is valid at least for
weak inelastic scattering which would indeed be
the case at low $T$ if only elastic scattering were present at $T=0$.
For elastic scattering, we could identify
\[ \rho \Sigma''(\omega\!=\!0,T\!=\!0)
\sim n_{i} \frac{S(S+1)}{\pi} \sin^{2} \phi_{0} 	,	\]
which would lead to
\begin{equation}
	\phi_{0} \sim \frac{\pi}{k} .		\label{p-shift}
\end{equation}
The resistivity due to the elastic scattering can
be calculated  by using (\ref{rate-f}) as the scattering rate.
To the leading order,
\begin{equation}
 \delta\rho_{e}^{\rm el}(T) =
\int^{\infty}_{0} \frac{d\omega}{2T \cosh^{2}\frac{\omega}{2T} }
 \frac{3\pi^{2} n_{i}}{4k^{2}n_{e}}  \rho_{e}^{(0)}
\left[ 1 -
A_{0} \left(\frac{\omega}{T_{K}}\right)^{\Delta}   \right] .
		\label{resis-el}
\end{equation}
Substituting (\ref{resist}), (\ref{p-shift}) and (\ref{resis-el})
into (\ref{resis-sum}), we would obtain  the inelastic
contribution to the
resistivity
$\delta\rho_{e}^{\rm in} < 0$ which certainly is impossible.
This contradiction indicates that  there are both
elastic and  inelastic  scattering
at $T=0$. For the two channel case, $k=2$, it has been suggested
that there is actually only inelastic scattering \cite{affl93}.

\section{Specific Heat and Entropy}	\label{SHaE}

To calculate the free energy, we can use the linked cluster theorem,
\begin{equation}
F=F_{0} -  \sum_{n=2}^{\infty} U_{n}  ,
			\label{free-en-exp}
\end{equation}
where $F_{0}$ is the free energy of the non-interacting Fermi sea.
The diagrams for the first three $U_{n}$ are drawn in
Figure~\ref{free-en-fig}.
The results after completing the Matsubara frequency summations
 are
\begin{eqnarray}
U_{2} &=& -\frac{3n_{i}}{16} g^{2} k \int
 d\epsilon \, d\epsilon_{1}
\frac{ f(\epsilon) - f(\epsilon_{1}) }
{\epsilon-\epsilon_{1}}
				\\
U_{3} &=& \frac{n_{i}}{8} g^{3} k  \int
  d\epsilon_{1} \, d\epsilon_{2} \, d\epsilon_{3}
\left[ \frac{ f(\epsilon_{1}) \, f(\epsilon_{2}) \, f(-\epsilon_{3})}
{(\epsilon_{3}-\epsilon_{1})(\epsilon_{3}-\epsilon_{2})}
+ {\rm 2 \;\; Permutations}  \right]			\label{u3}	\\
U_{4} &=&  \frac{3n_{i}}{64} g^{4} k^{2}  \int
 d\epsilon_{1} \, d\epsilon_{2}\,
d\epsilon_{3}\,  d\epsilon_{4}  \;
f(-\epsilon_{1}) \, f(\epsilon_{2}) \,
f(\epsilon_{3}) \,f(-\epsilon_{4})
				\nonumber  \\
 &  \times & \left[ \frac{1}{ (\epsilon_{1} - \epsilon_{2} )
(\epsilon_{3} - \epsilon_{4} ) }
\left( \frac{1-e^{\beta
( \epsilon_{2}-\epsilon_{1} + \epsilon_{3} - \epsilon_{4} ) } }
{ \epsilon_{1} - \epsilon_{2} - \epsilon_{3} + \epsilon_{4} }
+ \frac{e^{\beta(\epsilon_{2}-\epsilon_{1})}
-e^{\beta ( \epsilon_{3}-\epsilon_{4}) }}
{ \epsilon_{1} - \epsilon_{2} + \epsilon_{3} - \epsilon_{4} }
	\right) \right] .	\label{u4}
\end{eqnarray}
As noted by Kondo \cite{kond68},
it is quite delicate to extract the linear $T$ terms in (\ref{u3}) and
(\ref{u4}).
Here we follow the  method used by Kondo and present the calculations
only for the integrals not calculated in \cite{kond68}.

The calculation of $U_{2}$ is simple and straightforward.
The result is a constant plus $T^{2}$ corrections. To calculate
$U_{3}$, we define a new integral $I_{3}$,
\[  I_{3} = \int^{D}_{-D}
d\epsilon_{1} \, d\epsilon_{2} \, d\varepsilon_{3}
 \frac{ f(\epsilon_{1}) \, f(\epsilon_{2}) \, f(-\epsilon_{3})}
{( \epsilon_{3}-\epsilon_{1})_{\delta} \;
( \epsilon_{3}-\epsilon_{2})_{\delta}} ,  \]
where we have introduced the notation
\[ \frac{1}{(x)_{\delta}} = \frac{x}{x^{2} +\delta^{2}},
\hspace{.3in} \delta \rightarrow 0	.	\]
The $\delta \rightarrow 0$ limit is taken after
the integrations are completed.
Kondo showed \cite{kond68},
\begin{equation}
 U_{3} = 	\frac{3 n_{i}}{8} g^{3} k
\left( I_{3} - \frac{\pi^{2}}{6}T \right) ,		\label{u3-sep}
\end{equation}
so we  only sketch the calculation for $I_{3}$.
\[ I_{3} =  \int  d\epsilon_{1} \, d\epsilon_{2}
  f(\epsilon_{1}) \, f(\epsilon_{2}) \int^{D}_{-D} \frac{d\epsilon_{3}}
{ ( \epsilon_{3}-\epsilon_{1})_{\delta} \;
 (\epsilon_{3}-\epsilon_{2})_{\delta} }
- \int  d\epsilon_{1} \, d\epsilon_{2} \, d\varepsilon_{3}
 \frac{f(\epsilon_{1}) \, f(\epsilon_{2}) \, f(\epsilon_{3})}
{ ( \epsilon_{3}-\epsilon_{1})_{\delta} \;
 (\epsilon_{3}-\epsilon_{2})_{\delta} }	 .		 \]
The first integral has no linear $T$ contribution. Thus, up to
a constant term, we have
\begin{eqnarray*}
I_{3} &=&  -\frac{1}{3}  \int^{D}_{-D}
d\epsilon_{1} \, d\epsilon_{2}\, d\epsilon_{3} \;
 f(\epsilon_{1}) \, f(\epsilon_{2})   \, f(\epsilon_{3})	\\
 &  &  \times \left[
\frac{1}{ ( \epsilon_{3}-\epsilon_{1})_{\delta} \;
 (\epsilon_{3}-\epsilon_{2})_{\delta} }
+ \frac{1}{ ( \epsilon_{1}-\epsilon_{2})_{\delta} \;
 (\epsilon_{1}-\epsilon_{3})_{\delta} }
+ \frac{1} { ( \epsilon_{2}-\epsilon_{1})_{\delta} \;
 (\epsilon_{2}-\epsilon_{3})_{\delta} }
			 \right]  	 \\
 &=&  -\frac{1}{3}  \int^{D}_{-D}
  d\epsilon_{1} \, d\epsilon_{2} \, d\epsilon_{3}\;
\frac{ f(\epsilon_{1}) \, f(\epsilon_{2})   \, f(\epsilon_{3})	\;
[( \epsilon_{1}-\epsilon_{2})^{2}
+ ( \epsilon_{1}-\epsilon_{3}) \,
( \epsilon_{2}-\epsilon_{3}) ] \; \delta^{2} }
{ [( \epsilon_{1}-\epsilon_{2})^{2}+\delta^{2}] \,
[( \epsilon_{1}-\epsilon_{3})^{2}+\delta^{2}] \,
[( \epsilon_{2}-\epsilon_{3})^{2}+\delta^{2}] }   \\
&=&   -\frac{\pi^{2}}{3}  \int^{D}_{-D}
   d\epsilon_{1} \, d\epsilon_{2}\, d\epsilon_{3} \;
 f(\epsilon_{1}) \, f(\epsilon_{2})   \, f(\epsilon_{3})	\;
 \delta(\epsilon_{1}-\epsilon_{3}) \,
\delta(\epsilon_{2}-\epsilon_{3})  		\\
&=&  -\frac{\pi^{2}}{3}  \int^{D}_{-D}
   d\epsilon_{1} \;  [ f(\epsilon_{1})]^{3}
 = -\frac{\pi^{2}}{2} T  .
\end{eqnarray*}
Combining the  above result with (\ref{u3-sep}), we  obtain
\begin{equation}
 U_{3} = {\rm const.} - \frac{ n_{i} \pi^{2} }{4} g^{3}k \,  T  .
\label{u3-T}
\end{equation}
Next, let's turn to $U_{4}$ which  can be calculated in a similar way.
Defining a new function
\begin{eqnarray*}
 I_{4} &=&  \int
 d\epsilon_{1} \, d\epsilon_{2}\,
d\epsilon_{3}\,  d\epsilon_{4}  \;
f(-\epsilon_{1}) \, f(\epsilon_{2}) \,
f(\epsilon_{3}) \,f(-\epsilon_{4})			 \\
   &  \times & \left[ \frac{1}{ (\epsilon_{1} - \epsilon_{2} )_{\delta}
(\epsilon_{3} - \epsilon_{4} )_{\delta} }
\left( \frac{1-e^{\beta
( \epsilon_{2}-\epsilon_{1} + \epsilon_{3} - \epsilon_{4} ) } }
{ (\epsilon_{1} - \epsilon_{2} - \epsilon_{3} + \epsilon_{4})_{\delta} }
+ \frac{e^{\beta(\epsilon_{2}-\epsilon_{1})}
-e^{\beta ( \epsilon_{3}-\epsilon_{4}) }}
{ (\epsilon_{1} - \epsilon_{2} + \epsilon_{3} - \epsilon_{4})_{\delta} }
	\right) \right]					\\
 &=&  4  \int  d\epsilon_{1} \, d\epsilon_{2}\,
d\epsilon_{3}\,  d\epsilon_{4}  \;
\frac{f(-\epsilon_{1}) \, f(\epsilon_{2}) \,
f(\epsilon_{3}) \,f(-\epsilon_{4})	}
{ (\epsilon_{1} - \epsilon_{2} )_{\delta}
(\epsilon_{3} - \epsilon_{4} )_{\delta}
(\epsilon_{1} - \epsilon_{2} - \epsilon_{3} + \epsilon_{4})_{\delta} }  ,
\end{eqnarray*}
then following the same steps Kondo employed to prove (\ref{u3-sep}),
it's straightforward to show
\begin{equation}
U_{4}-	 \frac{3n_{i}}{64} g^{4} k^{2} I_{4}
 = \frac{3 \pi^{2} n_{i} }{32} g^{4} k^{2}  T .
\end{equation}
It can  be further verified that $I_{4}$ has no contribution
 linear  in  $T$. Therefore,
\begin{equation}
U_{4} = {\rm const.} + \frac{ 3 \pi^{2} n_{i} }{32} g^{4}k^{2} T  .
\label{u4-T}
\end{equation}
Substituting (\ref{u3-T}) and (\ref{u4-T}) into (\ref{free-en-exp})
and setting $n_{i}\!=\!1$ for simplicity,
we obtain the free energy shift due to the presence of the impurity
spin,
\begin{equation}
 F_{\rm imp}(T) = -E_{0} - T\ln 2 + \frac{\pi^{2}}{4}T
\left( kg^{3} -\frac{3}{8} k^{2}g^{4} \right)  +{\cal O}(g^{4}k) ,
		\label{fimp0}
\end{equation}
where $E_{0}$ is the ground state energy.

Since the free energy shift is RG invariant up to
an additive constant, we obtain
the scaling solution for the free energy shift
at $T\rightarrow 0$ after carrying out standard RG procedure,
\begin{eqnarray}
   F_{\rm imp}(T) &=& -E'_{0} - T\ln 2   +   \frac{\pi^{2}}{4}T
\left[ kg_{R}^{3}(T) -\frac{3}{8} k^{2}g_{R}^{4}(T) \right]
			\nonumber		\\
 & =&  -E'_{0} -T(\ln 2 - \frac{\pi^{2}}{2k^{2}})
-\frac{3\pi^{2}}{4} \zeta^{2} T  \left(\frac{T}{T_{K}}\right)^{2\Delta}
+{\cal O}[T(T/T_{K})^{3\Delta}]   ,
\end{eqnarray}
where $E'_{0}$  is in general different
 from $E_{0}$ of  (\ref{fimp0}).
The impurity specific heat is
\begin{equation}
C_{\rm imp} = -T \frac{\partial^{2} F_{\rm imp} }{\partial T^{2}}
= \frac{3\pi^{2}}{2}  \zeta^{2} \Delta
\left(\frac{T}{T_{K}}\right)^{2\Delta} .
			\label{s-heat}
\end{equation}
The critical exponent, $\alpha=2\Delta$, agrees with the previous
result \cite{andr84,affl91-1}.
The impurity entropy  is reduced to a universal value \cite{affl91-2}
\begin{equation}
 S_{\rm imp}(T=0) = \ln 2  -\frac{\pi^{2}}{2k^{2}} \simeq
\ln\left(2-\frac{\pi^{2}}{k^{2}}\right)    .  \label{entropy}
\end{equation}
Actually, the correction  to the entropy occurs
only when one first takes  the thermodynamic limit
${\cal N} \rightarrow \infty$  and then the limit $T \rightarrow 0$.
For a finite system,  the integrals in $U_{3}$ and $U_{4}$
should be replaced by  discrete  momentum
summations over the Brillouin zone which will
not give rise to contributions linear in $T$ to the free energy,
thus no correction to the bare $\ln 2$ term
of the entropy.
Therefore, the  finite system always remains doubly degenerate.

\section{Field Dependent Magnetization}		\label{FDM}

A natural question following the finite entropy
is whether or not there is a corresponding residual impurity
spin. To answer this question we calculate the field dependent
impurity magnetization defined as
the total
magnetization of the system subtracting the free Fermi sea
contribution.
In the presence of the external magnetic field, the leading order
diagrams to the free energy are shown in  Figure~\ref{free-en-Th}.
This is an infinite series.
The diagram with $l$ vertices has a contribution,
\begin{equation}
 \delta F_{l} =
-\frac{1}{\beta} \frac{k^{l}}{l} (-1)^{l+1}
\left[ \frac{-J}{{\cal N}\beta} \sum_{\vec{k},\omega_{n}} \sum_{\sigma=\pm}
\frac{-\sigma/2}
{i\omega_{n} - \epsilon_{\vec{k}} -\sigma h} \right]^{l}
 \sum_{\omega_{m},\sigma'=\pm} \left( \frac{-\sigma'/2}
{i\omega_{m}-i\omega_{0} -\sigma'  \mu_{S} h}  \right)^{l}   .
\end{equation}
The momentum summation over the conduction band is readily carried out,
\begin{equation}
\frac{J}{{\cal N}\beta} \sum_{\vec{k},\omega_{n}} \sum_{\sigma=\pm}
\frac{\sigma}
{i\omega_{n} - \epsilon_{\vec{k}} -\sigma h}
 = g \int^{D}_{-D} d\epsilon
\left[ f(\epsilon+h) - f(\epsilon-h) \right]
=- 2 g h .		\label{gh2}
\end{equation}
Summing  up all the diagrams in Figure~\ref{free-en-Th}  and
using (\ref{gh2}), the free energy shift
due to the impurity spin is
\[  F_{\rm imp}(T,h) =  \sum_{l}  \delta F_{l}
= \sum_{\sigma=\pm} \sum_{l=1}^{\infty}
\left(\frac{- \sigma gkh}{2}\right)^{l} \frac{1}{l !}
\left[ \frac{d^{l-1} f(z)}{dz^{l-1}}
		\right]_{z=i\omega_{0} + \sigma  \mu_{S} h}  . \]
Introducing the primitive of $f(z)$: $du(z)/dz=f(z)$, the series
can be summed up,
\[   F_{\rm imp}(T,h) =
\sum_{\sigma=\pm} \left[ u(i\omega_{0}+\sigma  \mu_{S} h - \sigma kgh/2)
-u(i\omega_{0} +\sigma \mu_{S} h) \right]		.  		 \]
The corresponding magnetization is
\[ \delta M(T,h) = - \frac{\partial F_{\rm imp}}{\partial h}
=  \left( \mu_{S} -\frac{kg}{2}\right)
 \tanh\left[\beta h\left(\mu_{S}-\frac{kg}{2}\right)\right]
-\tanh(\beta  \mu_{S} h)  .  \]
Adding to the above result the bare magnetization
of the impurity spin which just cancels
the second term of the above expression, we finally obtain
\begin{equation}
M(T,h) = \left(\mu_{S}-\frac{kg}{2}\right)
 \tanh\left[\beta h\left(\mu_{S}-\frac{kg}{2}\right)\right]  .
		\label{mTh}
\end{equation}
This is the leading order temperature and field dependent
magnetization.

In this section,
we shall mainly consider the case $\mu_{S}=1$, {\em i.e.} an equal
 gyromagnetic ratio.
The expression~(\ref{mTh}) has the form
characteristic  of
a  reduced free spin,
\[ M=S_{\rm eff} \tanh(S_{\rm eff} h/T).  \]
The fact that $S_{\rm eff} \rightarrow 0$ following a power law
implies  that the impurity spin is completely
screened  only for the infinite system. A finite system of size $L$
 behaves as if there is   a partially screened spin
$S_{\rm eff} \sim L^{-\Delta} $ since there is a minimum energy
unit $1/L$ to cut off the scaling.
It also suggests that the ground state degeneracy will
be lifted by  any small magnetic field at $T=0$.
Using the Maxwell relation
$\partial S/\partial h=\partial M/\partial T$,
the leading order magnetic field dependence of the entropy is
\begin{eqnarray}
S_{\rm imp}(T,h)
&=& \ln 2 + \int^{h}_{0} dh' \frac{\partial M(T,h')}{\partial T}
	\nonumber	\\
&=&  \ln 2 - \int_{0}^{\beta h |1-gk/2|}
 dx \frac{ x}{\cosh^{2}x}  \rightarrow 0,
\hspace{.1in} {\rm for} \; \; \left\{ \begin{array}{c}
		T \rightarrow 0  \\ h \neq 0  .
		\end{array}  \right.
\end{eqnarray}
We expect that this is   true to any order. In particular, the
entropy  change, $\pi^{2}/(2k^{2})$, in the next order (\ref{entropy}) will
be removed by the magnetic field.

{}From (\ref{mTh}), we obtain  the  low
temperature  magnetic susceptibility
and its scaling form \cite{andr84,affl91-1},
\begin{equation}
\chi_{\rm imp}(T) = \beta \left(1-\frac{kg}{2} \right)^{2}
= \beta \left(1-\frac{kg_{R}(T)}{2} \right)^{2}
\stackrel{T < T_{K}}{=} \left( \frac{k\zeta}{2} \right)^{2}
\frac{1}{T}  \left(\frac{T}{T_{K}}\right)^{2\Delta} ,
		\label{chi}
\end{equation}
indicating that the fixed point is a spin singlet
since $S_{\rm eff}^{2} \sim
T \, \chi_{\rm imp}(T) \sim T^{2\Delta} \rightarrow 0$ at $T\rightarrow 0$.
The spin is quenched very slowly compared with
the linear dependence in the  exactly screened case.
Also from (\ref{mTh}), we determine  the field dependence of
the zero temperature magnetization  \cite{affl91-1},
\begin{equation}
M(T=0,h) = 1- \frac{kg}{2}
= 1- \frac{kg_{R}(h)}{2}
 \stackrel{T < T_{K}}{=}
  \frac{k\zeta}{2} \left(\frac{h}{T_{K}}\right)^{\Delta} .
\end{equation}
This shows that the spin operator has the scaling dimension $\Delta$.

Using the  well known results for the
bulk specific heat and  magnetic susceptibility:
$C_{\rm bulk}=2 k \pi^{2}T\rho/3$ and
$\chi_{\rm bulk}=2k \rho$ respectively
\cite{ashc76},
the  leading order Wilson ratio
is determined from (\ref{s-heat}) and (\ref{chi}),
\begin{equation}
W=\frac{ \chi_{\rm imp}}{C_{\rm imp}}  \frac{C_{\rm bulk}}{\chi_{\rm bulk}}
= \frac{ \left( \frac{k\zeta}{2} \right)^{2}
\frac{1}{T}  \left(\frac{T}{T_{K}}\right)^{2\Delta} }
{ \frac{3\pi^{2}}{2}  \zeta^{2} \Delta
\left(\frac{T}{T_{K}}\right)^{2\Delta} }
\frac{ \frac{2}{3} k \pi^{2}T\rho }{ 2 k \rho } = \frac{k^{3}}{36} ,
\end{equation}
in agreement with the conformal field result \cite{affl91-1}.
The Wilson ratio is universal because the only parameter
having  a possible dependence on the cut-off scheme is $\zeta$
which is canceled out.

For $\mu_{S} \neq 1$, the external magnetic field
couples to  an unconserved  spin operator.
The magnetization
will acquire an anomalous dimension and
 we shall discuss  it in detail
in section~\ref{MSF}.

\section{Dynamical Susceptibility}		\label{DS}

For the one-channel Kondo problem,
it is well known that the impurity spin flipps at a
typical rate of Kondo temperature $T_{K}$.
At $T < T_{K}$, the impurity spin is effectively inert
leading to a fixed point of local Fermi liquid type
described by a phase shift.
To understand the multichannel case,
we calculate the dynamical
spin susceptibility. As we shall see, the typical spin flipping rate
is given by the temperature.

 We proceed as usual by
first calculating Matsubara spin-spin correlation function.
\begin{equation}
\chi_{f}(\tau)= \langle \hat{\rm T} \vec{S}(\tau) \cdot
\vec{S}(0) \rangle
= \frac{1}{\beta} \sum_{n} \chi_{f}(i\nu_{n}) \; e^{-i\nu_{n} \tau}  ,
\end{equation}
where $\nu_{n}= 2 n\pi/\beta$.
To the sub-leading order, the diagrams are shown in Figure~\ref{dyn-sus}.
The result of  calculating these diagrams is
\begin{equation}
\chi_{f}(i\nu_{n}) = \delta_{n,0} \frac{3\beta}{4}
\left[1-g^{2}k(\ln\beta D -\ln 2 +I_{0})\right]
-\frac{3g^{2}k}{4} K(i\nu_{n})  	,		\label{chi-f}
\end{equation}
where $ I_{0} = \int^{\infty}_{0} dx
[ \tanh (x/2)/x - 1/(1+x) ] \simeq 0.125 $
and
\begin{eqnarray}
K(i\nu_{n}) &=& \int^{D}_{-D}\int^{D}_{-D}
 \frac{d\varepsilon_{1} \, d\varepsilon_{2}}
{(\varepsilon_{1}-\varepsilon_{2})^{2} + \nu_{n}^{2} }
\left[ \frac{ f(\varepsilon_{1}) - f(\varepsilon_{2}) }
{\varepsilon_{1}-\varepsilon_{2} }
- f'(\varepsilon_{1}) \delta_{n,0} \right]
		\nonumber 		\\
 &=&   - \frac{ \pi}{ | \nu_{n} | } ( 1- \delta_{n,0})
+ {\cal O}(D^{-1})	.		 \label{K}
\end{eqnarray}
To  understand the result we just derived, let's
consider $\chi(z)$ in
the complex frequency plane $ i\nu_{n} \rightarrow z$
outside a disk of radius of order $T$ (Figure~\ref{z-plane}).
Knowing its values at a set of discrete points
on the imaginary axis, $z = i\nu_{n}$, is enough to give
a unique analytic continuation,
\begin{equation}
\chi_{f}(i\nu_{n}) = \frac{3\pi}{4} g^{2} k \frac{1}{|\nu_{n}|}
\rightarrow  \frac{3\pi}{4} g^{2} k  \; \frac{i \; {\rm sgn(Im }z) }{z  } ,
\end{equation}
where sgn(Im $z$) means  taking the sigh of the imaginary part of  $z$.
When $z$ approaches the real axis,
$ z \!=\! \omega+i \, \delta$,
$\chi_{f}(\omega+i\delta)=\chi_{f}'(\omega) + i \;
{\rm sgn}(\delta) \; \chi_{f}''(\omega)$,
and we find
\begin{equation}
\chi_{f}''(\omega) = \frac{3\pi}{4} g^{2} k \;  \frac{1}{\omega} .
		\label{chif-ome}
\end{equation}
This is valid in the whole plane outside a disk
of radius of order $T$. There is no energy scale to cut off
$1/\omega$ dependence. Certainly for $\omega \ll T$,
this dependence should be flattened out as seen
from the $\nu_{n}\!=\!0$ expression of (\ref{chi-f}).
If we recall that in the exactly screened
 Kondo problem($k=1$),
 $\chi_{f}''(\omega)/\omega \sim 1/(\omega^{2}+T_{K}^{2})$,
the system gives sizable magnetic response
only when  all the contributions up to the energy scale $T_{K}$
are included. In other words, the impurity spin is effectively quenched
on time scale longer than $1/T_{K}$.
In the multichannel case, the role of $T_{K}$ is played by
the temperature $T$!
The impurity is only marginally quenched
since the time scale is $1/T$.

In the other limit $\omega \ll T$, carrying out the
analytic continuation as in \cite{gan93-1},  we find
\begin{equation}
\frac{ \chi_{f}''(\omega, T) }{\omega } \sim
T^{2 \Delta - 1}  .		\label{t1t}
\end{equation}
Note that if $\mu_{S} \gg 1$, the coupling of
 the nuclear magnetic moment to
the impurity
spin dominates and
 the above result is proportional to the
NMR relaxation rate $1/(T_{1}T)$.
In the limit $k \gg 1$, local probe basically sees
a nearly free impurity spin. In the two-channel case,
$k\!=\!2$, $2\Delta\!=\!1$,
we see that
(\ref{t1t}) will have no power law divergence in $T$
except possible logarithmic dependence.

In the two-channel case,
 numerical work in the non-crossing
approximation \cite{cox88}
and the solution at a special Toulouse point \cite{emer92}
have found
$\chi_{f}''(\omega) \sim \tanh(\omega/2T) /(\omega^{2}+T_{K}^{2})$.
The $\tanh(\omega/2T)$ term has  also been produced by
the conformal field theory method  and is a property
of the fixed point itself. It would be
interesting to see if  the
remaining Lorentzian form with width $T_{K}$ can be
reproduced from the perturbation of
the leading irrelevant operator in  the exact conformal field theory
calculations.
For $k > 2$, we do not expect a Lorentzian form with finite width $T_{K}$
for the following reason.
The multiplicative renormalization  factor
for $\chi_{f}$(see next section) will bring an anomalous
dimension $\omega^{2\Delta}$ in (\ref{chif-ome}).
We see that $\chi_{f}''(\omega) \sim \omega^{2\Delta -1}$
for $T < \omega < T_{K}$. Since $2\Delta=4/(k+2) < 1$,
this frequency dependence
 cannot be reconciled with a Lorentzian form with a finite width.

\section{Magnetic Susceptibility For $\mu_{S} \neq 1$ }	  \label{MSF}

Besides the time ordered impurity spin-spin correlation
function studied in the last section,
we can also define
\begin{eqnarray}
\chi_{fe}(\vec{q},\tau) &=& \langle \hat{\rm T}
\vec{S}_{e}(\vec{q},\tau) \cdot \vec{S}(\tau\!=\!0) \rangle
= \frac{1}{\beta} \sum_{n} \chi_{fe}(\vec{q},i\nu_{n})
\; e^{-i\nu_{n} \tau} 			,		\\
\chi_{e}(\vec{q},\tau) &=& \langle \hat{\rm T}
\vec{S}_{e}(\vec{q},\tau) \cdot \vec{S}_{e}(0,0) \rangle
= \frac{1}{\beta} \sum_{n} \chi_{e}(\vec{q},i\nu_{n})
\; e^{-i\nu_{n} \tau} 			,
\end{eqnarray}
where the conduction electron spin operator is defined as
\begin{equation}
\vec{S}_{e}(\vec{q})= \frac{1}{2} \sum_{\vec{k}}
\sum_{ \stackrel{ \mu,\nu=\pm }{ \lambda = 1 \cdots k} }
c^{\dagger}_{\vec{k},\mu\lambda} \vec{\sigma}_{\mu\nu}
c_{\vec{k}+\vec{q},\nu\lambda} .
\end{equation}
The Feynman diagrams
for these two correlation functions are shown
in Figure~\ref{dyn-sus} and calculated
in the appendix~\ref{dyn-correl}.

When the impurity spin has a different gyromagnetic
ratio from  the conduction electrons,
a uniform  external magnetic field couples to an unconserved spin
operator $\vec{S}_{h} \!=\!
\vec{S}_{e}(\vec{q}\!=\!0)+\mu_{S}\,\vec{S}$.
The magnetic susceptibility is then only  RG invariant
up to a multiplicative renormalization factor,
\begin{equation}
\chi_{\rm imp}(T, g, D) = \left[ Z_{h}(g,g_{R}(D')) \right]^{2}
\chi_{\rm imp}(T, g_{R}(D'), D')  ,
		\label{zs-def}
\end{equation}
where $Z_{h}$ is the multiplicative renormalization factor
for the operator $\vec{S}_{h}$
and we recall $g_{R}(D)\!=\!g$. If $\mu_{S}=1$, then
$Z_{h} \equiv 1$.
The RG equation for $\chi$ is
\begin{eqnarray}
 & & \left[ D \frac{\partial}{\partial D}
+ \beta(g)  \frac{\partial}{\partial g}
+ 2 \gamma_{h}(g) \right] \chi_{\rm imp}(T, g, D) = 0  ,		\\
 & & \gamma_{h}(g) = D \frac{\partial  \ln Z_{h} }{\partial D} .
\end{eqnarray}
The perturbative result for $\chi$ to the sub-leading
order can be found from the results
for three dynamical correlation functions calculated
in the last section and appendix~\ref{dyn-correl},
\begin{eqnarray}
 & & \chi_{\rm imp}(T, g, D) = \frac{4}{3}
\left[ \chi_{e}(\vec{q}\!=\!0,\nu_{n}\!=\!0)
+ 2 \mu_{S} \chi_{fe}(\vec{q}\!=\!0,\nu_{n}\!=\!0)
+ \mu_{S}^{2} \chi_{f}(\nu_{n}\!=\!0)  \right]    \nonumber   \\
 & & \hspace{.3in}
= \beta \left\{ \left(\mu_{S} - \frac{gk}{2} \right)^{2}
\left[ 1 - g^{2}k ( \ln \beta D -\ln 2 +I_{0} ) \right]
- g^{2}k \left(\mu_{S} - \frac{gk}{2} \right) \ln 2 \right\} .
	\hspace{.1in}		\label{chi-mus}
\end{eqnarray}
The leading term of $\gamma_{h}(g)$ can be obtained
from (\ref{chi-mus}) using the same method
of section~\ref{RGEaS},
\begin{equation}
\gamma_{h}(g) = \frac{\mu_{S} -1}
{ 2 \mu_{S} - gk  } \; g^{2} k  .
\end{equation}
Its value at the fixed point, $\gamma_{h}(g^{*})$,
gives the anomalous dimension of the operator
$\vec{S}_{h}$.
The anomalous dimensions for the operators $\vec{S}$ and
$\vec{S}_{e}(\vec{q}\!=\!0)$ are the limiting values
at $\mu_{S} \rightarrow \infty$ and $\mu_{S} \rightarrow 0$
respectively.

To obtain the scaling solution, we first find
the multiplicative renormalization factor,
\begin{equation}
Z_{h}(g,g_{R}(D')) = \int^{g_{R}(D')}_{g} dg'
\frac{\gamma_{h}(g')}{\beta(g')}
= \left( \frac{g k - 2 \mu_{S}}{gk -2 } \right)^{2}
\left( \frac{g_{R}(D') k - 2 }{g_{R}(D') k -2  \mu_{S} } \right)^{2} .
		\label{zs-f}
\end{equation}
Setting $D'\!=\!T$ in (\ref{zs-def}) and (\ref{zs-f}),
we obtain the leading order susceptibility(since we only
have leading order $Z_{h}$),
\begin{equation}
\chi_{\rm imp}(T, g, D)
= \left( \frac{g k - 2 \mu_{S}}{gk -2 } \right)^{2} \;
\left[ \chi_{\rm imp}(T, g, D) \right]_{\mu_{S}=1}  .
\end{equation}
This is the main result of this section: the static magnetic
response is the same,
up to a constant factor,  no matter
whether or not the conduction electrons
and the impurity spin have the same gyromagnetic ratio.
I believe this conclusion remains true for all overscreened
cases and even for the exactly
screened case of $k=1$.
If the initial starting point belongs to weak coupling, $g \rightarrow 0$,
\begin{equation}
\chi_{\rm imp}(T, g, D) = \mu_{S}^{2} \;
\left[	\chi_{\rm imp}(T, g, D) \right]_{\mu_{S}=1}	.
\end{equation}
This is the well known result for all kind of Kondo problems:
the contribution to the magnetization from the conduction electrons
is suppressed by a factor $T_{K}/D$ compared with  that of the
 impurity spin \cite{lowe84}.
Extending the treatment in this section to the field dependent
magnetization and dynamical susceptibilities is straightforward.

\section{Knight Shift}			\label{KS}

Knight shift  gives direct measurement of the
spatial structure of the conduction electron screening cloud.
Consider an impurity spin sitting at the origin and a
 uniform magnetic field $h$
being applied to the system, Knight shift measures
the magnetization in the space surrounding the impurity spin,
\begin{equation}
M(\vec{r})  =  \sum_{ \stackrel{\sigma=\pm}{\lambda=1 \cdots k} }
\sigma \;
 \langle c^{\dagger}_{\sigma,\lambda}(\vec{r}) \;
c_{\sigma,\lambda}(\vec{r}) \rangle
= \chi(\vec{r}) \;  h  ,
 \hspace{.3in}
\chi(\vec{r}) = \frac{1}{\cal N} \sum_{\vec{q}}
\chi(\vec{q}) \; e^{i\vec{q} \cdot \vec{r} } .
\end{equation}
We implicitly
assume that the free Fermi sea contribution to $M(\vec{r})$
has been subtracted.
 The result for the Knight shift is contained in the
two general dynamical susceptibilities defined in the last section,
\begin{equation}
\chi(\vec{q}) =  \frac{4}{3} \left[ \chi_{e}(\vec{q},i\nu_{n}\!=\!0)
+ \mu_{S}  \chi_{fe}(\vec{q},i\nu_{n}\!=\!0) 	\right]  .
\end{equation}
Using the results from the appendix~\ref{dyn-correl}, we find
to the sub-leading order in $1/k$,
\begin{eqnarray}
\chi(\vec{q}) &=& - \frac{gk}{2T}  \Pi_{0}(\vec{q}) \left\{
\left(\mu_{S} - \frac{gk}{2} \right)
\left[ 1 - g^{2}k ( \ln \frac{ D}{T} - \ln 2 + I_{0} ) \right]
- \frac{\ln 2}{2} g^{2}k \right\}
		\nonumber 	\\
 & & - \frac{\ln 2}{2T} g^{2} k
\left(\mu_{S} - \frac{gk}{2} \right)	\Pi_{1}(\vec{q}) ,   \\
\Pi_{0}(\vec{q}) &=& - \frac{1}{\rho {\cal N}}
\sum_{\vec{k}}
\frac{ f(\epsilon_{\vec{k}}) - f(\epsilon_{\vec{k}+\vec{q}}) }
{ \epsilon_{\vec{k}}  - \epsilon_{\vec{k}+\vec{q}} }     ,	 \\
\Pi_{1}(\vec{q}) &=& \frac{1}{2\ln2} \frac{1}{\rho^{2} {\cal N}^{2}}
\sum_{\vec{k},\vec{k}'}
\frac{ \tanh\frac{\epsilon_{\vec{k}} }{2T} }
{ \epsilon_{\vec{k}} - \epsilon_{\vec{k}'} }  \left[
\frac{ f(\epsilon_{\vec{k}}) - f(\epsilon_{\vec{k}'+\vec{q}}) }
{ \epsilon_{\vec{k}}  - \epsilon_{\vec{k}'+\vec{q}} }
-  \frac{ f(\epsilon_{\vec{k}'}) - f(\epsilon_{\vec{k}'+\vec{q}}) }
{ \epsilon_{\vec{k}'}  - \epsilon_{\vec{k}'+\vec{q}} }   \right] ,
\end{eqnarray}
where $I_{0} \simeq 0.125$,  given in the last section.
We have included proper normalization factors in the definitions
of $\Pi_{0}(\vec{q})$ and $\Pi_{1}(\vec{q})$
such that $\Pi_{0}(\vec{q}\!=\!0)=1$ and $\Pi_{1}(\vec{q}\!=\!0)=1$.
Note that the Fourier transform of $\Pi_{0}(\vec{q})$
has the well known Friedel oscillation form in space \cite{rude54}.
The Fourier transform of $\Pi_{1}(\vec{q})$ should give  similar
spatial variation. The key point is that
their Fourier transforms only depend on $k_{F} r$, where $r$
is the distance to the impurity spin.

 Since $\chi(\vec{q})$ satisfies the
 RG equation,
\begin{equation}
 \left[ D \frac{\partial}{\partial D}
+ \beta(g)  \frac{\partial}{\partial g}
+  \gamma_{h}(g)  + [\gamma_{h}(g)]_{\mu_{S}=0} \right]
\chi(\vec{q}) = 0	.
\end{equation}
The scaling solution can be found
in a similar way to the last section,
\begin{equation}
\chi(\vec{q},T,g,D ) =  Z_{h}(g,g_{R}(T))
 \left[ Z_{h}(g,g_{R}(T)) \right]_{\mu_{S}=0}
\chi(\vec{q},T,g_{R}(T),T )	.
\end{equation}
Using the results for $Z_{h}$ and $g_{R}$,
we obtain  to the leading order,
\begin{equation}
\chi(\vec{q},T,g,D )  =  \frac{g k(gk - 2 \mu_{S})}{(gk -2)^{2} }
\chi_{\rm imp}(T) \, \Pi_{0}(\vec{q})  ,		\label{chiq}
\end{equation}
where $\chi_{\rm imp}(T)$ is the uniform susceptibility found
in section~\ref{FDM}.
A knowledge of $Z_{h}$ to the sub-leading order could
extend (\ref{chiq}) to the next order.
When Fourier transforming (\ref{chiq})
to the real space,  $M(\vec{r})$ is a function
of $k_{F} r$.
So the only length scale is $1/k_{F}$.
Thinking over why the length scale  $v_{F}/T_{K}$
does not appear in $\chi(\vec{q})$, we observe that
  $\Pi_{0}$ and $\Pi_{1}$
are functions of $q/k_{F}$.
A new length scale $v_{F}/T_{K}$ would appear only
if the logarithmic singularity were cut off by
$v_{F}q$  in the limit $v_{F}q > T$.
This doesn't happen!
It means that the multichannel system is critical only
in the time  direction,  not in the space.

We can extend the statement that there is only one length
scale $1/k_{F}$ to the one-channel problem.
In this case,
the effective Kondo interaction flows to strong coupling
at low energy.
{}From the renormalization group point of view,
it means that summing up leading(or leading plus sub-leading)
logarithmic series is not enough.
But  reorganizing the perturbation expansion
in the coupling constant $g$ into successive logarithmic
series(the first term of each series has successive higher power in $g$)
 according to
 the renormalization group
is still possible.
This is equivalent to write
$\beta(g)$ as an expansion in $g$.
The occurrence of a strong  coupling fixed point
means that we need all terms in the expansion of
$\beta(g)$ to describe the low energy behavior.
Suppose we could find all of them(infinite terms),
we would have a correct low energy theory by summing
up all terms.
The fact that  the infrared
singularity is not cut off by $v_{F} q$
does {\em not} change merely because we have to include
high order logarithmic series.
The necessary consequence is that $v_{F}/T_{K}$
will never have the chance to appear as a length scale.
Certainly, if the flowing of an effective interaction
to strong coupling leads to a phase transition like in
higher dimensions, the above argument may not hold.
But the impurity problem has a dimension  $0+1$,
prohibiting a phase transition.
Thus, our calculation provides an explicit analytical
demonstration that the
screening cloud has a spatial size
of order $1/k_{F}$, and  agrees with
 the well known experimental result \cite{boyc74}.

\section{Discussion}

We have carried out a comprehensive study on
the non-Fermi liquid fixed point of the multichannel Kondo model
emphasizing several intriguing aspects.
As the actual realization of this model in heavy fermion
and metallic glassy systems is  still an open question, a thorough
understanding of the model should be very helpful in
devising new experimental tests and making comparison with experimental
results.
Part of the driving force behind the recent resurgence of interest
is due to the resemblance of its low energy behavior
to the normal state properties of high Tc cuprates.
This similarity  has been further exploited recently \cite{emer93-1}
arguing that a situation similar to the two-channel
model is realized in the copper-oxide plane.

One interesting  followup subject is to study the
possibility of converting the local non-Fermi liquid
fixed point into a coherent  lattice one  by forming
an overscreened Kondo lattice, especially in two or three
dimensions.
The first step to study  the two-channel-two-impurity problem
has been carried out using the numerical renormalization
group method \cite{inge92}.
The one-impurity fixed point is found unstable against
developing correlations between the
impurity spins. This is expected
in the view of the asymptotical screening
of the impurity spin.
One artificial way to suppress  magnetic correlations between
impurity spins is to go to infinite dimension.
Unfortunately, the presence of  finite
degeneracy on each site prevents development of
true coherence. In any realistic situation,
nontrivial magnetic correlations
must intervene to lift the residual degeneracy of the
impurity fixed point.
If one wishes to follow the successful route of the
heavy fermion theory  from impurity to lattice once again,
in the current intensive search of non-Fermi liquid models
to describe the normal state of high Tc cuprates,
only a non-degenerate fixed
point of an impurity model has the chance to succeed.
Development along this direction in search  of a promising
starting point  of impurity model has been reported
recently \cite{giam93}.
It is worthwhile mentioning that a quantum critical
system with a Fermi surface has a true coherent
non-Fermi liquid  infrared fixed point in high dimensions. An example
is an electron gas coupled to transverse
gauge field where a self-consistent solution of
the infrared fixed point has been derived \cite{gan93-2}.
It is very interesting to note that this system is
quantum critical at $T\!=\!0$ in $2D$ and $3D$
but without developing long range order.

\acknowledgments
The author acknowledges useful conversations with
D. Cox, N. Prokof'ev,
P. Stamp, A. Tsvelick and C. Varma,   and especially
his indebtedness to I. Affleck,
N. Andrei and P. Coleman whose patient explanation
has been indispensable to the author's understanding of this
subject. This work was supported in part by NSERC and CIAR of Canada.

\appendix{ Feynman Rules }	\label{feyn-rule}

To be self-contained,
we list the rules for constructing Feynman diagrams
in the presence of  an external
magnetic field. These rules  also define our convention.
\begin{enumerate}
	\item For the contributions of
the $n$th order in the Kondo interaction,
${\cal O}(J^{n})$, draw all topologically distinct diagrams
 with $n$ vertices.
	\item For each conduction electron propagator, draw a solid
line,
\begin{equation}
\hspace{-1.5in}
\begin{picture}(80,20)
\thicklines
\put (10,0){\vector(2,0){30}}
\put (30,0){\line(2,0){35}}
\put (25,7){$\vec{k},i\omega_{n}$}
\end{picture}
= \; G^{(0)}_{\mu\nu}(\vec{k},i\omega_{n}) \;
= \; \frac{-\delta_{\mu\nu}}{i\omega_{n}-\epsilon_{\vec{k}}-\mu h}.
\end{equation}
There is a summation over each momentum $\vec{k}$ of
the internal conduction electron propagator.
For each pseudofermion propagator,  draw a dashed line,
\begin{equation}
\hspace{-1.5in}
\begin{picture}(80,20)
\thicklines
\put (10,0){\line(2,0){7}}
\put (22,0){\line(2,0){7}}
\put (34,0){\vector(2,0){7}}
\put (39,0){\line(2,0){7}}
\put (51,0){\line(2,0){7}}
\put (63,0){\line(2,0){7}}
\put (25,7){$\vec{k},i\omega_{n}$}
\end{picture}
= \; {\cal G}^{(0)}_{\mu\nu}(i\omega_{n}) \;
= \; \frac{-\delta_{\mu\nu}}{i\omega_{n}-i\omega_{0}-\mu \mu_{S} h}.
\end{equation}
	\item Each vertex is associated with a  factor,
\begin{equation}
\hspace{-1in}
\begin{picture}(280,90)(10,10)
\thicklines
\put(50,40){\circle*{10}}
\put (10,10){\vector(4,3){25}}
\put (30,25){\line(4,3){20}}
	\put (50,40){\vector(-4,3){25}}
	\put (30,55){\line(-4,3){20}}
\put (50,40){\line(4,3){10}}
\put (66,52){\vector(4,3){10}}
\put (82,64){\line(4,3){10}}
	\put (90,10){\line(-4,3){10}}
	\put (74,22){\vector(-4,3){10}}
	\put (58,34){\line(-4,3){10}}
\put (2,0){$\vec{k},i\omega_{n},\mu$}
\put (2,76){$\vec{k}', i\omega_{n}', \mu'$}
\put (94,0){$i\omega_{m},\nu$}
\put (94,76){$i\omega_{m}', \nu'$}
	\put(150,45){ $ = -\frac{J}{4{\cal N}\beta} \;
\vec{\sigma}_{\mu'\mu} \cdot \vec{\sigma}_{\nu'\nu} \;
\delta_{\omega_{n}+\omega_{m},\omega_{n}'+\omega_{m}'} . $ }
\end{picture}
\end{equation}
	\item Each independent internal frequency is summed over.
	\item  Each conduction electron loop contributes
a factor $-k$ and each pseudofermion loop contributes a factor $-1$.
	\item For a diagram of order $J^{n}$, there is a numerical
factor: (Number of different connections)/$n!$. This combinatorial
factor will be given explicitly in the figures of this paper for the
diagrams we calculate.
\end{enumerate}

\appendix{Dynamical Spin-Spin Correlation Functions}	\label{dyn-correl}

Three dynamical correlation functions have been defined in the
sections~\ref{DS} and \ref{MSF}.
To the order ${\cal O}(1)+{\cal O}(1/k)$,
their diagrams  are shown in  Figure~\ref{dyn-sus}.
Each diagram contains two external vertices
which are represented by  two open ends at the left and
right sides.
The external vertex at each end is joined by two
propagators.
If the spin indices of these two propagators are
$\sigma_{1}$ and $\sigma_{2}$, the corresponding
vertex is associated with a vector of matrices
$\vec{s}_{\sigma_{2}\sigma_{1}}$.
The two vectors of matrices from the
two external vertices at the two ends form a scalar product
$ \vec{s}_{\sigma_{2}\sigma_{1}} \cdot
\vec{s}_{\sigma_{4}\sigma_{3}} $.
 In $\chi_{e}$, only one of the two external vertices
 has the two joining
conduction electron propagators differing in their
momenta by $\vec{q}$.

The result for $\chi_{f}(i\nu_{n})$ is given in (\ref{chi-f}).
The results for the other two are
\begin{eqnarray}
\chi_{fe}(\vec{q},i\nu_{n}) &=&
- \delta_{n,0} \; \frac{3}{8} g k \beta \left\{ \Pi_{0}(\vec{q})
\left[1-g^{2}k(\ln\beta D -\ln 2 +I_{0})\right]
+  g \ln 2  \; \Pi_{1}(\vec{q})  \right\}
	\nonumber		\\
 & & 	+ (1 - \delta_{n,0}) \chi_{fe}(\vec{q},i\nu_{n}) ,   \\
\chi_{e}(\vec{q},i\nu_{n}) &=&
- \frac{g k}{2} \, \delta_{n,0} \; \chi_{fe}(\vec{q},i\nu_{n})
+  \frac{3\ln 2}{16} g^{3} k^{2} \beta   \Pi_{0}(\vec{q})
- (1 - \delta_{n,0}) \chi_{fe}(\vec{q},i\nu_{n})   ,
\end{eqnarray}
where $\Pi_{0}$ and $\Pi_{1}$ are defined in
section~\ref{KS} and for $\nu_{n} \neq 0$,
\begin{eqnarray}
\chi_{fe}(\vec{q},i\nu_{n})
&=&   \frac{3}{8} g^{2} k \left[
gk \, \Pi_{0}(\vec{q}) K(i\nu_{n})
+ 2  K(\vec{q},i\nu_{n}) - 2 L(\vec{q},i\nu_{n}) \right] , \\
K(\vec{q},i\nu_{n}) &=&
 \frac{1}{\rho^{2} {\cal N}^{2}}
\sum_{\vec{k},\vec{k}'}
\frac{ 1 }{ \epsilon_{\vec{k}} - \epsilon_{\vec{k}'+\vec{q}} }
\frac{ f(\epsilon_{\vec{k}}) - f(\epsilon_{\vec{k}'}) }
{  (\epsilon_{\vec{k}}  - \epsilon_{\vec{k}'} )^{2} + \nu_{n}^{2} } , \\
L(\vec{q},i\nu_{n}) &=&  \frac{1}{\rho^{2} {\cal N}^{2}}
\sum_{\vec{k},\vec{k}'}
\frac{ 1 }{ \epsilon_{\vec{k}} - \epsilon_{\vec{k}'} }
\frac{ f(\epsilon_{\vec{k}'}) - f(\epsilon_{\vec{k}'+\vec{q}}) }
{  ( \epsilon_{\vec{k}'}  - \epsilon_{\vec{k}'+\vec{q}} )^{2}
		+ \nu_{n}^{2} }  .
\end{eqnarray}
Note that $K(\vec{q}\!=\!0,i\nu_{n})=K(i\nu_{n})$, is a generalization
of $K(i\nu_{n})$ to finite $\vec{q}$.
In the expressions for $\chi_{fe}$ and $\chi_{e}$, we have dropped
contributions of order $1/D$
to the $\nu_{n}\!=\!0$ components with respect to $\beta$.

\bibliography{hfmag,hightc}

\figure{Feynman diagrams for the conduction electron self-energy.
			\label{self-en-fig}	}

\figure{Diagrams for the free energy. The solid line represents the bare
conduction electron propagator. The big circles stand for self-energies.
			\label{free-en-fig}	}

\figure{Diagrams for the free energy in the magnetic field.
			\label{free-en-Th}	}

\figure{Diagrams for three
 dynamical spin correlation functions.
			\label{dyn-sus}	}

\figure{ Analytic continuation in the complex frequency plane
carried out in the region outside a disk of radius of order $T$.
		\label{z-plane}	}

\end{document}